\documentclass[pra,twocolumn,a4paper,superscriptaddress]{quantumarticle}
\usepackage{graphicx,amsmath,bm,color}
\usepackage{latexsym}
\usepackage{amssymb}
\usepackage{amsfonts}
\usepackage[english]{babel}
\usepackage[bbgreekl]{mathbbol}
\usepackage{hyperref}
\usepackage{subfigure}
\usepackage{braket}
\usepackage{color}

\newcommand{\ketbra}[2]{|#1\rangle\hspace{-2.4pt}\langle #2| }

\begin{document}

\title{Proposal for witnessing non-classical light with the human eye}
\author{A. Dodel}
\affiliation{Department of Physics, University of Basel, Klingelbergstrasse 82, 4056 Basel, Switzerland}
\author{A. Mayinda}
\affiliation{Department of Physics, University of Basel, Klingelbergstrasse 82, 4056 Basel, Switzerland}
\author{E. Oudot}
\affiliation{Department of Physics, University of Basel, Klingelbergstrasse 82, 4056 Basel, Switzerland}
\author{A. Martin}
\affiliation{Group of Applied Physics, University of Geneva, Ch. de Pinchat 22, 1211 Geneva, Switzerland}
\author{P. Sekatski}
\affiliation{Institut for Theoretische Physik, Universitat of Innsbruck, Technikerstra{\ss}e 25, A-6020 Innsbruck, Austria}
\author{J.-D. Bancal}
\affiliation{Department of Physics, University of Basel, Klingelbergstrasse 82, 4056 Basel, Switzerland}
\author{N. Sangouard}
\affiliation{Department of Physics, University of Basel, Klingelbergstrasse 82, 4056 Basel, Switzerland}

\date{\today}

\begin{abstract}
We give a complete proposal showing how to detect the non-classical nature of photonic states with naked eyes as detectors. The enabling technology is a sub-Poissonian photonic state that is obtained from single photons, displacement operations in phase space and basic non-photon-number-resolving detectors. We present a detailed statistical analysis of our proposal including imperfect photon creation and detection and a realistic model of the human eye. We conclude that a few tens of hours are sufficient to certify non-classical light with the human eye with a p-value of 10\%.
\end{abstract}
\maketitle

\section{Introduction \& motivations} 
Efforts have been recently devoted to the realization of quantum experiments with the human eye. This endeavor is however challenging. The proposal of Ref. \cite{Brunner08} which uses many entangled photon pairs to realize a Bell test with the eye does not allow one to violate a Bell inequality with a realistic model of the eye.  Refs. \cite{Sekatski09,Sekatski10} which propose to amplify entanglement of a photon pair through a phase covariant cloning, can lead to entanglement detection with eye-based detectors provided that strong assumptions are made on the source. While no assumption is needed on the functioning of the eye, it is necessary to assume that the source produces true single photons. From a practical point of view, phase-covariant cloning is also difficult to implement. In particular, cloning is inherently multimode when implemented with a non-linear crystal as suggested in Ref. \cite{Sekatski09}. The undesired modes can be filtered out but at the price of introducing substantial loss. Ref. \cite{Caprara16} provides a technically simpler solution by using displacement operations on single-photon entanglement. This proposal allows one to detect entanglement with the eye without assumption on the source but needs a precise description of the visual system. Indeed, the entanglement witness proposed in Ref. \cite{Caprara16} relies on a well-defined model of the eye thus requiring a detailed characterization of the human eye. Importantly, in both Ref. \cite{Sekatski09} and \cite{Caprara16}, entanglement is detected before the amplification. That is, these proposals allow one to conclude that few-photon entanglement can be detected by the human eye upgraded by phase-covariant cloning and displacement operations respectively. The question we address in this manuscript is how the quantum nature of light can be \textit{directly} detected with the eye.

The motivations are twofold. First, our proposal is a fascinating attempt to get closer to the quantum world. Indeed, it is conceptually very different from standard quantum optics experiments where measurements are done by photon detectors and the sole role of experimentalists in the measurement process is to analyse the experimental data stored on a computer. The envisioned experiment is unitary until the eye, so if a collapse happens it does not happen before the eye. Second, such an experiment interfaces quantum light and biological systems. Inspired by the great success of quantum optics in revolutionizing communications \cite{Gisin07}, metrology \cite{Giovannetti11}, sensing \cite{Degen16} or computing \cite{Ladd10}, this experiment of a new kind may flourish with important applications for biomedical research. \\

As stated before, the proposal of Ref. \cite{Caprara16} is appealing as it uses simple ingredients, namely single-photon entanglement and displacement operations. In this manuscript, we derive a witness for non-classical states and we show how the same ingredients allow one to reveal the non-classical nature of a superposition state with the eye. Our witness needs no assumption on the photon number produced by the source or on the precise modelling of the eye. It simply relies on the assumption that the probability to detect light increases with the photon number. While entanglement detection requires measurements in different bases, the experiment that we propose is simpler as it uses displacement operations with fixed amplitudes and phases. It does not need interferometric stabilization of optical paths and is very robust against loss. We show, through a detailed feasibility study including a realistic model of the human eye with a reasonable recovery time as well as imperfect photon creation and detection, that a few tens of hours are sufficient for our witness to conclude about non-classicality with a p-value of 10\%. Our results point towards a concrete proposal for implementing the first experiment where the quantum nature of light is revealed directly with the human eye.\\

\section{Witnessing non-classicality with rudimentary detectors} 

Coherent states $\ket{\alpha}$ of a harmonic oscillator (or a mode of the electromagnetic field) saturate the uncertainty relations for any pair of quadratures as well as for amplitude and phase \cite{MilburnWalls}. In addition, they are eigenstates of the positive frequency part of the quantized field and vector potential operators \cite{Glauber}.
%If a mode of the quantized electromagnetic field is in a coherent state $\ket{\alpha}$ then the expectation values of the corresponding operators reproduce the values and evolution of classical field (before quantization).  In addition, coherent states saturate the uncertainty relations for any two field quadratures as well as for amplitude and phase. 
For these reasons, the set of coherent states is thought as the most classical subset of all possible pure states of light. In this context, a state which can be expressed as a mixture of coherent states $\ket{\alpha}$
\begin{equation}
\label{classical}
\rho_{\text{class}} = \int \mathrm{d}^2\alpha\-\ p(\alpha) \ketbra{\alpha}{\alpha}, \-\ \text{with} \-\ p(\alpha) \geq 0
\end{equation}
is considered classical, and any state which cannot be decomposed in this way is then non-classical. It is easy to see that the convex combination of coherent states in Eq.~\eqref{classical} satisfies $\frac{\langle \hat N^2 \rangle- \langle \hat N \rangle}{\langle \hat N \rangle^2} \geq 1$ with $\hat N$ the number operator~\cite{comment}. Hence, a photon-counting detector can be used to witness the non-classical nature of a light state. If the photon-counting results reveal $\frac{\langle \hat N^2 \rangle- \langle \hat N \rangle}{\langle \hat N \rangle^2} < 1,$ we can indeed conclude that the measured state is non-classical. Note that all non-classical states lead to entanglement when combined with the vacuum on a beamsplitter \cite{Asboth05}. The link with entanglement helps clarifying the notion of non-classical states. \\

Moreover for few photon states, $\langle \hat N^2 \rangle- \langle \hat N \rangle$ can be approximated by $\sim 2 \langle \ketbra{2}{2}\rangle$ and $\langle \hat N \rangle^2$ by $\sim \langle \ketbra{1}{1} \rangle^2.$ Hence, one can use a 50/50 beamsplitter and two non-photon-number-resolving detectors to witness the non-classical nature of few photon states by checking that the two-fold coincidences ($\sim \langle \ketbra{2}{2} \rangle/2$) are smaller than the product of singles ($\sim \langle \ketbra{1}{1} \rangle^2 /4$), cf. \cite{Sekatski12} for a proper derivation. Can one still use this criterion in presence of other kinds of detectors? We now address the question of the conditions required to witness the non-classical nature of a light source with a 50/50 beamsplitter and two detectors.\\

Let us consider an arbitrary detector with a binary outcome, one corresponding to click, the other one to no-click. We label $p_{s}(\alpha)$ the probability to get a click when a coherent state $\ket{\alpha}$ impinges on such a detector. In a scenario where two of these detectors are placed after a 50/50 beamsplitter, the probability to get a twofold coincidence with any classical state is given by 
$p_{c}(\rho_{\text{class}})=\int \mathrm{d}^2\alpha\-\ p(\alpha) p_s(\alpha/\sqrt{2})^2$ whereas the probability of a single detection is given by $p_{s}(\rho_{\text{class}})=\int \mathrm{d}^2\alpha\-\ p(\alpha) p_s(\alpha/\sqrt{2})$. This simply comes from the fact that a coherent state splits into two similar coherent states on a beamsplitter $\ket{\alpha}\overset{\text{BS}}{\longrightarrow} \ket{\frac{\alpha}{\sqrt{2}}}_t\otimes\ket{\frac{\alpha}{\sqrt{2}}}_r$.  The Cauchy-Schwarz inequality $\int f(\mu)^2 \mathrm{d}\mu \int g(\mu)^2 \mathrm{d}\mu \geq \left(\int f(\mu) g(\mu) \mathrm{d}\mu\right)^2$ for $f=1$, $g=p_{s}(\alpha/\sqrt{2})$ and $\mathrm{d}\mu=p(\alpha)\mathrm{d}^2\alpha $ then implies
\begin{equation}
\frac{p_{c}(\rho_{\text{class}})}{p_{s}(\rho_{\text{class}})^2} \geq 1.
\end{equation}
In other words, any detector can be used to witness non-classicality as long as one has two copies of this particular detector. It suffices to place these detectors after a 50/50 beamsplitter and to record the number of singles and coincidences. If the ratio between the probability of having a coincidence and the square of the probability of singles is smaller than one, we can safely conclude that the measured state is non-classical. We show in the appendix A that the ratio between the coincidence and the product of singles is a witness for non-classicality even if the two detectors after the beamsplitter are not identical and the beamsplitter is not balanced, as long as $p_s(\alpha)$ is an increasing function of the photon number $|\alpha|^2$ for both detectors. These results are used in the next section to show how to detect non-classical states with the human eye.\\

\section{Witnessing non-classicality with the human eye}
Let us start this section by recalling how to model the response of the human eye to weak light stimuli. In a landmark experiment Hecht, Shlaer and Pirenne tested the capability of the human eye to detect light pulses containing only a few photons \cite{Hecht42}, see also  \cite{Rieke98}. In their experiment, an observer was presented with a series of multimode thermal light pulses and asked to report when the light is seen. Similar results have been obtained much more recently with coherent light pulses (monomode light also having a Poissonian photon-number distribution) \cite{Tinsley16}, thus indicating that the response of the eye does not depend on the number of modes. Interestingly, the results of both experiments are very well reproduced by a model in which coherent states are sent onto a threshold detector preceded by loss. In particular, the experimental data of Ref. \cite{Hecht42} is compatible with a threshold at $\theta=7$ photons and an efficiency of $\eta_e=8\%,$ see Fig. 1 in Ref. \cite{Caprara16}. Note that these numbers depend on the psychophysics, \textit{i.e.} the dark adaptation, the choice of dead-times and methods for eliciting responses from the observer about his experience of light stimuli. In particular, the recent results reported in Ref. \cite{Tinsley16} are compatible with lower thresholds and several references \cite{Donner91,Field05} suggest higher efficiencies. In the remainder of the paper, we keep the model of the eye with parameters associated to the seminal work of Hecht and co-workers ($\theta=7$ and $\eta_e=8\%$). We show that these parameters are conservative, \textit{i.e.} higher efficiencies or lower thresholds reduce the number of experimental runs that are needed to conclude about non-classicality. \\ 

\begin{figure}
  \centering
\includegraphics[width=0.8\columnwidth]{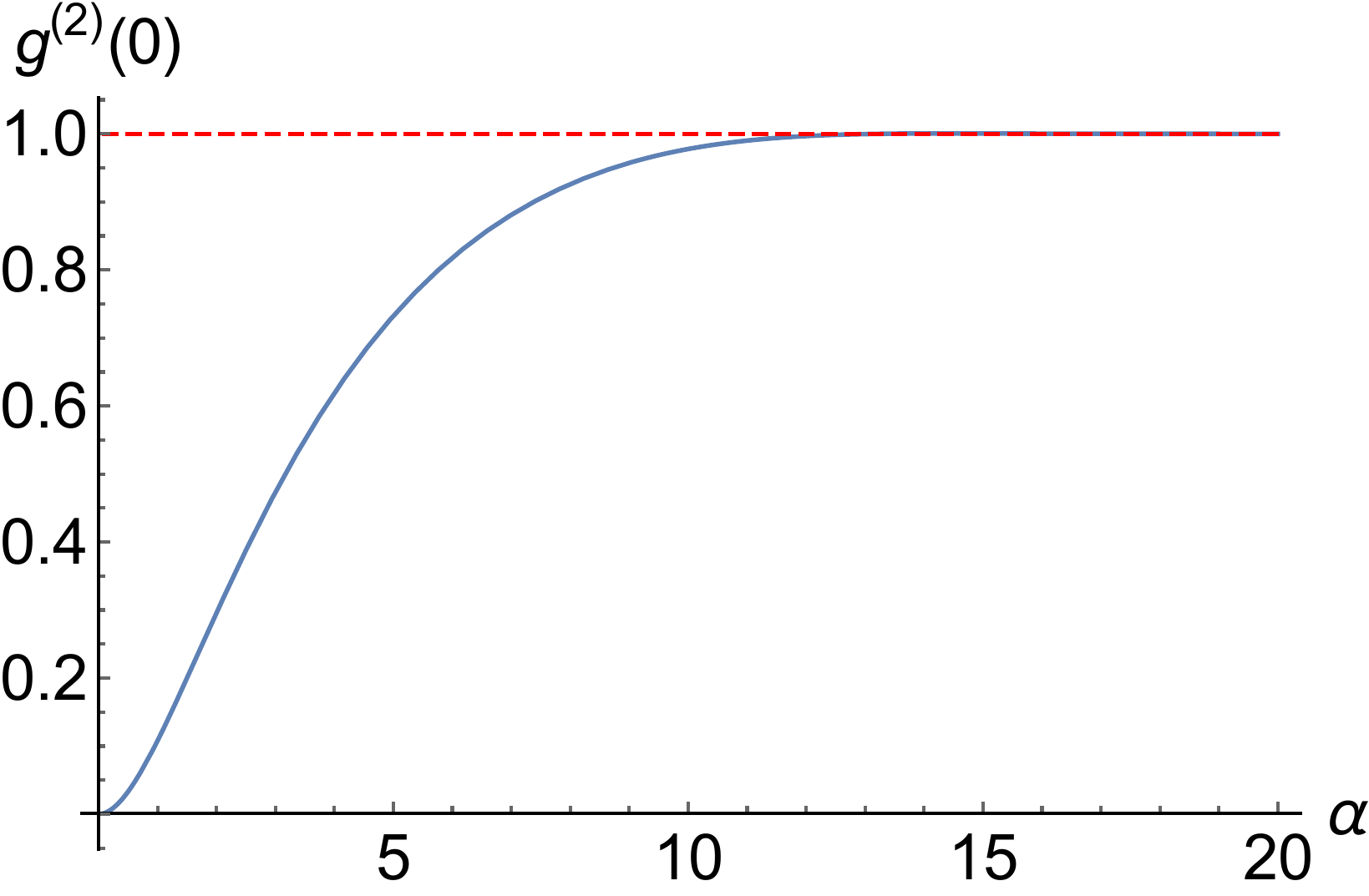}
\caption{Result of an auto-correlation $(g^{(2)}(0))$ measurement in which two eyes are placed after a 50/50 beamsplitter. The ratio between the probability to see light with both eyes and the square of the probability to see light with one eye is recorded for an input state $\mathcal{D(\alpha)}(\ket{0}+\ket{1})/\sqrt{2},$ considering real $\alpha.$ We here show this ratio as a function of $\alpha.$ Ratios smaller than one (red dashed line) witness the non-classical nature of the state.} 
\label{Fig1}
\end{figure} 

Given the witness for non-classical states presented in the previous section, we envision an experiment where two eyes are placed after a beamsplitter. The event ``click" corresponds to the case where the observer sees light, ``no-click" where no light is seen. The experiment is repeated several times to access the probability to see light with one of the two eyes as well as the joint probability to see light with both eyes. The ratio between the coincidences and the product of singles is then used to reveal non-classicality. This ratio is labelled $g^{(2)}(0)$ in analogy to the standard autocorrelation measurement. \\

To make a complete proposal, we still need to find a quantum state for which the non-classical nature can be revealed in such a setup. Note that sub-Poissonian states, \textit{i.e.}\ states for which the distribution in photon-number space is narrower than the one of a coherent state with the same mean photon number, are natural candidates for achieving $g^{(2)}(0) <1$ with threshold detectors such as the human eye. This is because there is a regime where, for the same probability of singles, the narrow photon-number distribution of a sub-Poissonian state yields a lower coincidence probability than the one of the corresponding coherent state. As an illustration, consider an ideal threshold detector and a Fock state that has enough photons to eventually make one of the detectors click, but not enough to give a coincidence.\\
While Fock states with large photon numbers are challenging to produce, a sub-Poissonian state can be obtained in practice by displacing a superposition of vacuum and single-photon Fock state in phase space. The resulting state $\mathcal{D}(\alpha)\left|\frac{1}{\sqrt{2}}\left(0+1\right)\right\rangle$, where $\mathcal{D}(\alpha)$ stands for a displacement operation, indeed has a variance in photon-number space that is $\frac{1+8\vert\alpha\vert^2-4Re(\alpha)^2}{2+4\vert\alpha\vert^2+4Re(\alpha)}$ times that of a coherent state with the same mean photon number. This ratio admits values that are below one, and interestingly, for a given strength of the displacement $\vert\alpha\vert^2$, it is minimal and always inferior to unity when $\alpha$ is real. Consequently, from here on we will only consider real displacements.\\

Fig. \ref{Fig1} shows the value of $g^{(2)}(0)$ obtained when sending such a state on a 50/50 beamsplitter followed by two eyes as a function of the amplitude of $\alpha.$ We see that the non-classical nature of $\mathcal{D}(\alpha)\left|\frac{1}{\sqrt{2}}\left(0+1\right)\right\rangle$ can be detected with the human eye as long $\alpha \leq 13.3.$ For larger $\alpha,$ the two eyes always see light and the ratio between coincidences and singles tends to one. However, in the range of displacement values $\alpha\sim 10,$ one can expect non-negligible occurrence frequency for the event ``seen" for both eyes. These encouraging estimations compel us to make a detailed feasibility study, \textit{i.e.}\ to propose a practical way to create a single photon superposed with vacuum, to account for imperfect generation efficiency, channel loss, limited detection efficiencies and to conclude about the statistics that is required to witness non-classicality with the human eye. \\

\section{Proposed experiment}
\begin{figure}
  \centering
\includegraphics[width=1\columnwidth]{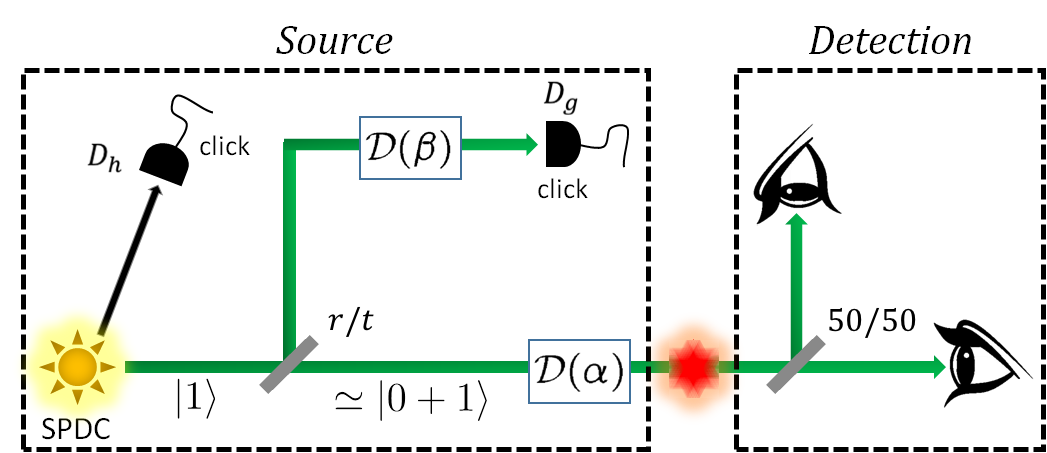}
\caption{Schematic representation of the experiment envisioned to witness the non-classical nature of a superposition state $\mathcal{D}(\alpha)(\ket{0}_t+\ket{1}_t)/\sqrt{2}$ with the human eye. The superposition $(\ket{0}_t+\ket{1}_t)/\sqrt{2}$ is prepared by first sending a single photon into an unbalanced beamsplitter and by subsequent detection of the reflected mode with a photon detector preceded by a displacement operation. For displacements with a small enough amplitude, this projects the transmitted mode into a state close to the desired superposition. This superposition state is then displaced to produce the non-classical state of interest. A 50/50 beamsplitter and two eyes are then used to analyse this state with a measurement analogous to an auto-correlation measurement.} 
\label{Fig2}
\end{figure}

The experiment we envision is shown in Fig. \ref{Fig2}. A source based on spontaneous parametric down-conversion is used to create photon pairs, the detection (on detector $D_h$ in Fig. \ref{Fig2}) of one photon from a given pair serving to herald the presence of its twin. The latter is then sent into a $50/50$ beamsplitter to create path-entanglement, \textit{i.e.}\ entanglement of the form $(\ket{0}_t\ket{1}_r - \ket{1}_t\ket{0}_r)/\sqrt{2}$ between the transmitted and reflected modes of the beamsplitter which share a single photon. The reflected mode is subsequently detected with a non-photon-number-resolving detector (detector $D_g$ in Fig. \ref{Fig2}) preceded by a displacement in phase space $\mathcal{D}(\beta)$. With the appropriate displacement amplitude, such a measurement performs a pretty good measurement along the $x$ direction of the Bloch sphere having $\ket{0}$ and $\ket{1}$ as its north and south pole respectively \cite{Caprara15a}. In other words, with the appropriate displacement, a detection click projects the transmitted mode into a state close to $(\ket{0}_t+\ket{1}_t)/\sqrt{2}.$ Such a state is then displaced in phase space, split using a 50/50 beamsplitter and sent to human observers. The single and coincidence events are recorded and the experiment is repeated until the observers can conclude about the non-classical nature of the superposition state with enough statistical confidence. As it is not clear what psychophysical test would allow to distinguish a dim flash of light occurring in the left vs. the right eye and a temporal discrimination with a single observer would require unrealistic delays, we envision an experiment with two observers, each reporting on whether he/she sees light each time a detection click is obtained on $D_g.$ We show below how to get a triggering rate compatible with a synchronization of the two observers' answers. \\

Note that in this setup, one can tune the transmission coefficient of the first beamsplitter along with the displacement amplitude $\beta$, effectively modifying the input state for the autocorrelation measurement. Finally, we observed that the closest state to $\mathcal{D}(\alpha)\left|\frac{1}{\sqrt{2}}\left(0+1\right)\right\rangle$ is obtained by choosing a highly unbalanced beamsplitter with transmission $t \sim 1$ and using a displacement $\mathcal{D}(\beta)$ with almost zero amplitude. In this case, we get a very partially entangled state and maximum coherence of the conditional state $(\ket{0}_t+\ket{1}_t)/\sqrt{2}$ is restored by measuring the reflected mode almost along the $z$ direction and post-selecting the case where a click is obtained.  This favors larger fidelities of the conditional state because the measurement noise is reduced when it gets closer to the z direction \cite{Caprara15a}. However, the probability to get a click drops when the transmission of the beamsplitter increases. There is thus a trade-off between the ``quality'' of the states produced by the source and the rate at which they are produced. The parameters $\beta$ and $t$ have to be optimized in view of the statistics needed to witness non-classicality, cf. below.\\

Several requirements need to be satisfied for implementing the experiment proposed in Fig. \ref{Fig2}. (i) The efficient generation of pure, indistinguishable and narrowband single photons is the first one. A straightforward way to create photons with these properties from spontaneous parametric down-conversion is to combine short, Fourier-limited pump pulses with a narrow-band filtering of the heralding photons. This results in Fourier-limited heralded photons with the spectrum of the pump \cite{URen05}. To ensure a high coupling efficiency of these heralded photons into an optical fiber, a plane wave pump is required and the heralding photons need to be spatially filtered with a single mode fiber before being detected. This projects the heralded photons into the fundamental spatial mode of the fiber, and hence allows one to reach very high coupling efficiencies \cite{Guerreiro13}. 
(ii) The photons need to have a color that can be seen by the human eye. This can be fulfilled with a pump at $405$nm down-converted into non-degenerate photon pairs at $1536$ and $550$nm. The advantage is threefold. $550$nm is very well suited for the human eye and the photons in the telecom band can be efficiently filtered both spatially and in frequency. The telecom mode can also be seeded with a stable cw telecom laser to generate the coherent states that are needed for the displacement operations, cf. below. 
(iii) The click rate on the detector $D_g$ in Fig. \ref{Fig2} needs to be adapted to the timescale of the response of the human eye as it sets a start for the observers. This can be done by reducing the repetition rate of the pump laser with an optical chopper. The heralding rate on $D_h$ and thus on $D_g,$ can then be easily set by tuning the laser intensity and the duty cycle of the optical chopper, c.f. below. (iv) To implement the displacement operations, we need an unbalanced beamsplitter and coherent pulses with Poissonian photon distribution that are indistinguishable from the photons at 550 nm in all degrees of freedom. This can be done using difference frequency generation. More precisely, we propose to use a second non-linear crystal, identical to the first one and pumped by the same laser but with a narrow seed of the telecom mode. In contrast to spontaneous parametric down-conversion, the seed results in coherent states at 550\,nm with the characteristics of the pump laser, \textit{i.e.}\ Fourier-limited coherent states with the spectrum of the pump \cite{Bruno13}. Since the coherent states created in this way and the single photons at 550\,nm are generated from the same pump, their indistinguishability is insensitive to the pump fluctuations. Note also that with a ps pump, the effect of frequency fluctuations of the telecom laser is negligible. The slow fluctuations in intensity of the latter can be recorded and taken into account once the measurements are done. In the worst case, they can be monitored and corrected with a feedback loop. Albeit with different wavelengths, the proposed technique has already been used successfully in various experiments \cite{Bruno13,Monteiro15}.\\

\begin{figure}
  \centering
\includegraphics[width=1\columnwidth]{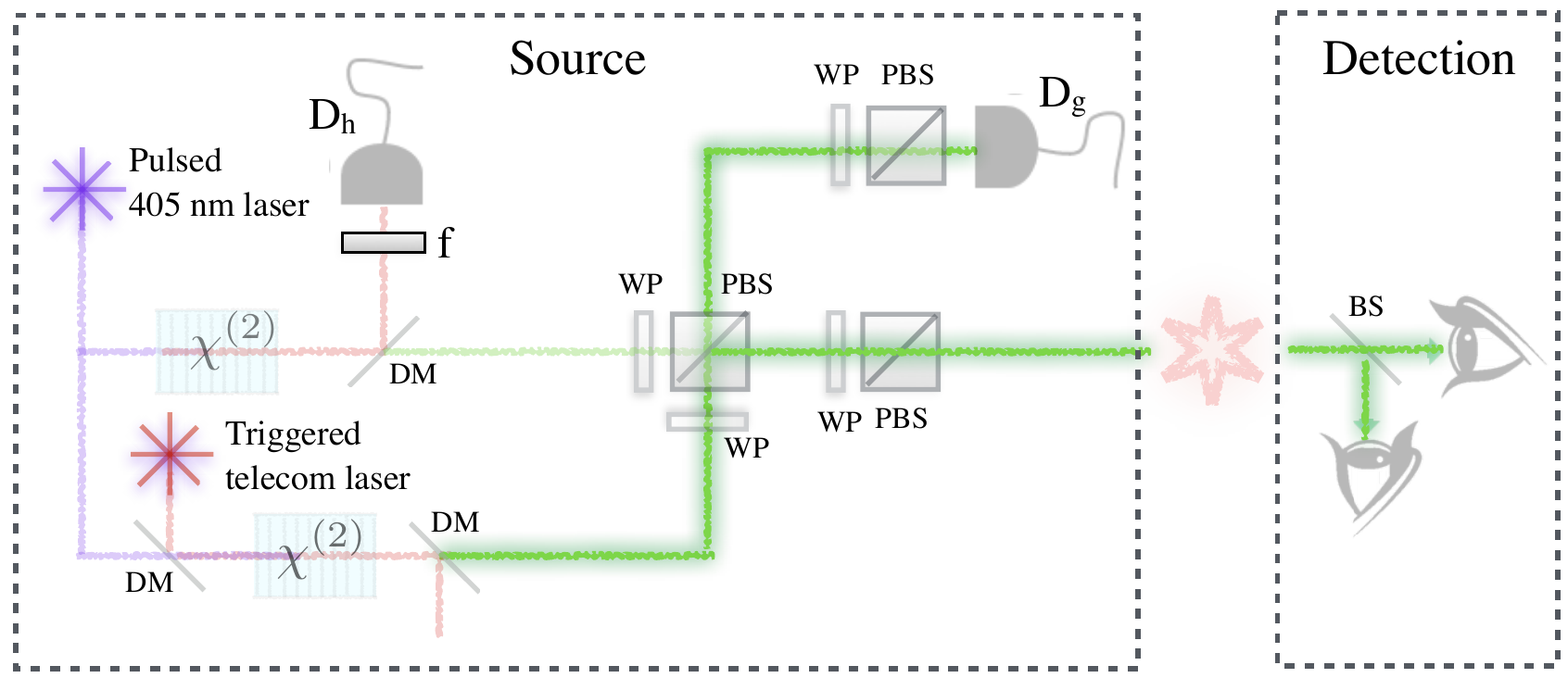}
\caption{Schematic of the setup to produce superposition states close to $\mathcal{D}(\alpha)(\ket{0}_t+\ket{1}_t)/\sqrt{2}$ and to detect their quantum nature with the human eye. Star: laser,  $\chi^2:$ non-linear crystal, DM: dichroic mirror, f: filters, WP: wave-plates, PBS: polarizing beamsplitter, BS: beamsplitter. See text for details.} 
\label{Fig2bis}
\end{figure}

Concretely, we envision an experiment where a Ti-Sa laser is doubled to create $2-3$ps pulses at $405$nm with a repetition rate of $80$ MHz, see Fig. \ref{Fig2bis}. These pulses are then used to pump two crystals in order to be down-converted to $1536$ and $550$nm respectively. The first crystal will be used to create pure single photons at $550$nm by picking up a single spatial and frequency mode of the photons at $1536$nm with a monomode fiber and a narrowband spectral filter. Coherent states that are indistinguishable from the photons at $550$ nm are generated by seeding the second crystal with a pulsed telecom laser. Let us emphasize that the critical point of this experimental implementation is the noise. In standard experiment, the noise is filtered out by analyzing the detection times to discriminate between true and false events. As the response of the human eye is not fast enough for such a temporal discrimination, we need to be sure that a limited number of undesired photons can reach the eye of the observer. First, we propose to decrease the repetition rate of the pump laser to 1.6 MHz using an optical chopper with a duty cycle of 0.02. By tuning the pump intensity to get a pair emission probability of $0.8\times10^{-3}$ and considering a global detection efficiency of 0.08 for $D_h$ (\textit{i.e.} a coupling efficiency of 0.8, a filter transmission of 0.4 and a raw detection efficiency of 0.25), we get a heralding rate on $D_h$ of $\sim 100$ Hz. Moreover, we consider a coupling efficiency of the heralded photon at $550$nm of $\eta_c=0.8$ in agreement with the experimental results reported e.g. in Ref. \cite{Guerreiro13}. The detection efficiency of the visible detector in the upper arm of Fig. \ref{Fig2} is assumed to be $\eta_d=0.5$ which is realistic even when including the transmission loss from the source to the detector and the inefficiencies of linear optical elements. We neglect  mismatches in the indistinguishability of the photons and coherent states at $550$nm, which is well justified given the results of Ref. \cite{Bruno13} where the visibility of the Hong-Ou-Mandel interference between a single photon and a coherent state created via identical crystals as described before was only limited by the statistics of the coherent state. We set the transmission $t = 98\%$ which, together with the value of the displacement $\beta \sim 0.08$ chosen to minimize the total number of experimental runs (cf. below), ensures that $1\%$ of the heralds on $D_h$ lead to a click on $D_g$. Meanwhile the conditional state generated on the lower arm shows a near maximal $95\%$ fidelity with respect to $\mathcal{D}(\alpha)\ket{\frac{1}{\sqrt{2}}(0+1)}$.\\

The dominant noise in this scenario comes from the coherent states that are used for the displacement operations. We propose to trigger the seed that is used to generate these coherent states on detections in $D_h$. In this case, the noise is $\sim 100$ times greater than the signal. To reduce it further, a pulse picker is placed in front of the eyes which is triggered by detections on $D_g.$ Considering an extinction ratio of 1:2000, we get a signal-to-noise ratio of $\sim 20,$ which should be more than enough to perform the proposed measurement. Note that the pulse picker also filters out other sources of noise, including the spontaneous emission of the crystal used to generate single photons at 550 nm (that is negligible with respect to the noise due to coherent states). Note also that $\sim 100$ns are needed to trigger the pulse picker on detections by $D_g,$ which requires a delay line of 20m of fiber, representing negligible loss for typical attenuation $<12$dB/km at $550$nm.

\section{Statistics}
To conclude the feasibility analysis of the proposed experiment, we now turn to the question of statistics, and determine the number of runs needed to exclude the possibility that the observed finite statistics are the result of measurements on a classical state. This is a particularly relevant question in our case, as the repetition rates that can be attained with the human eye are much lower than the slowest commercial detectors. The statistical study that we describe in this section aims at estimating the time-resource that an experimenter would have to allocate to such an experiment for the efficiencies discussed in the previous section, depending on the accuracy he wants to achieve.\\

The statistical issue is essentially an estimation of the odds of having $g^{(2)}(0)<1$ from a classical photon-number distribution. To answer this we consider the multinomial joint probability
\begin{equation}
\begin{split}
P(N_s,N_c) = &p_c^{N_{c}}(p_s-p_c)^{N_s-N_c}(1-p_s)^{N-N_s}\\
&\times{N \choose N_c,N_s-N_c,N-N_s}
\end{split}
\end{equation}
of obtaining $N_s$ singles and $N_c$ coincidences out of $N$ experimental runs, from the knowledge of the single and coincidence probabilities in one round $\{p_{s},p_{c}\}$. Note that we assume here that the single probability on each eye is identical, and that the runs are independently and identically distributed (i.i.d.). Further note that the form of the above distribution, whose natural variables are $N_c$ and $N_s-N_c$, stresses the dependence of the events ``single" and ``coincidence". Indeed we have defined a single in one arm regardless of the situation in the other arm, hence a coincidence is counted as a single as well. The outcome ``single only" has an occurrence probability $p_s-p_c$ as can be seen in the multinomial expression. Both the quantum scenario presented before, with $\{p_s(\rho_{\text{q}}), p_c(\rho_{\text{q}})\}$ depending on the non-classical state $\rho_\text{q}$, and the classical one with $\{p_s(\rho_{\text{c}}), p_c(\rho_{\text{c}})\}$ such that $p_c(\rho_{\text{c}})\geq p_s^2(\rho_{\text{c}})$ give rise to a probability distribution that we label respectively by $P^q(N_s,N_c)$ and $P^c(N_s,N_c).$ \\

We then choose an estimator $\chi$ which is a function of the total number of singles $N_s$ and coincidences $N_c$ observed in $N$ rounds of the experiment, cf. below. For a given $N$, this estimator takes the value $\chi(N_s, N_c)$ with probabilities $P^q(N_s,N_c)$ and $P^c(N_s,N_c)$ in the quantum and classical scenarios. The probability of observing a value of $\chi$ smaller than a given value $\chi_0$ in the quantum (classical) case after $N$ rounds is thus given by
\begin{equation}
P(\chi^{q/c} \leq \chi_0) = \sum_{N_s,N_c|\chi(N_s,N_c)\leq \chi_0} P^{q/c}(N_s,N_c).
\end{equation}
On one side, the quantum distribution tells us what is the probability with which we can expect to observe (in a quantum experiment) a value of $\chi$ smaller or equal to some value $\chi_0$. We write this probability
\begin{equation}
\label{define_pstop}
P_\text{stop}=P(\chi^q\leq\chi_0).
\end{equation}
On the other side, the classical distribution allows us to define the p-value $\epsilon$ associated with the rejection of the null hypothesis ``the state is classical" once a value $\chi_0$ is observed. This p-value is given by
\begin{equation}
\label{define_chi0}
\epsilon = \max_{p_c \geq p_s^2} P(\chi^c \leq \chi_0)
\end{equation}
where the maximum is taken over all classical scenarios satisfying $p_c(\rho_{\text{c}}) \geq p_s^2(\rho_{\text{c}}).$ Alternatively, we can read the relation \eqref{define_chi0} as a definition of the critical value of the estimator $\chi_0$ which needs to be obtained in order to rule out all classical states with a confidence of $1-\epsilon$. %Note that the quantum distribution is not relevant once the experiment is performed: only the value of the estimator $\chi$ actually observed matters. This distribution is only useful to study what can be expected from the experiment before it is built, and how likely we are to observe a significant violation with it. 
Choosing first the p-value, Eq. \eqref{define_chi0} gives $\chi_0$ which can then be used to get the probability to stop at the $N^{th}$ run using Eq. \eqref{define_pstop}. The average number of runs that is needed to rule out classical states can finally be estimated as  (cf. Appendix B)
\begin{equation}
\langle N \rangle \simeq \sum_{j\geq 0} \frac{n(2j+1)}{2} \left(P_\text{stop}(n(j+1)) - P_\text{stop}(nj)\right)
\end{equation}
where $n$ is a coarse-graining parameter used to make the computation faster. \\

% We compute the value $\chi_0$ such that the probability that the classical distribution $\bar \chi^{c}$ satisfies $P(\bar \chi^{c}\leq\chi_0)\leq\epsilon$ and we deduce the relevant probability $P_{\text{stop}}=P(\bar \chi^{q}\leq\chi_0).$ The latter is the probability that after $N$ experimental runs, one gets the required statistics to exclude the classical distribution within probability $1-\epsilon.$\\

\begin{figure}
  \centering
\includegraphics[width=0.6\columnwidth, keepaspectratio=true]{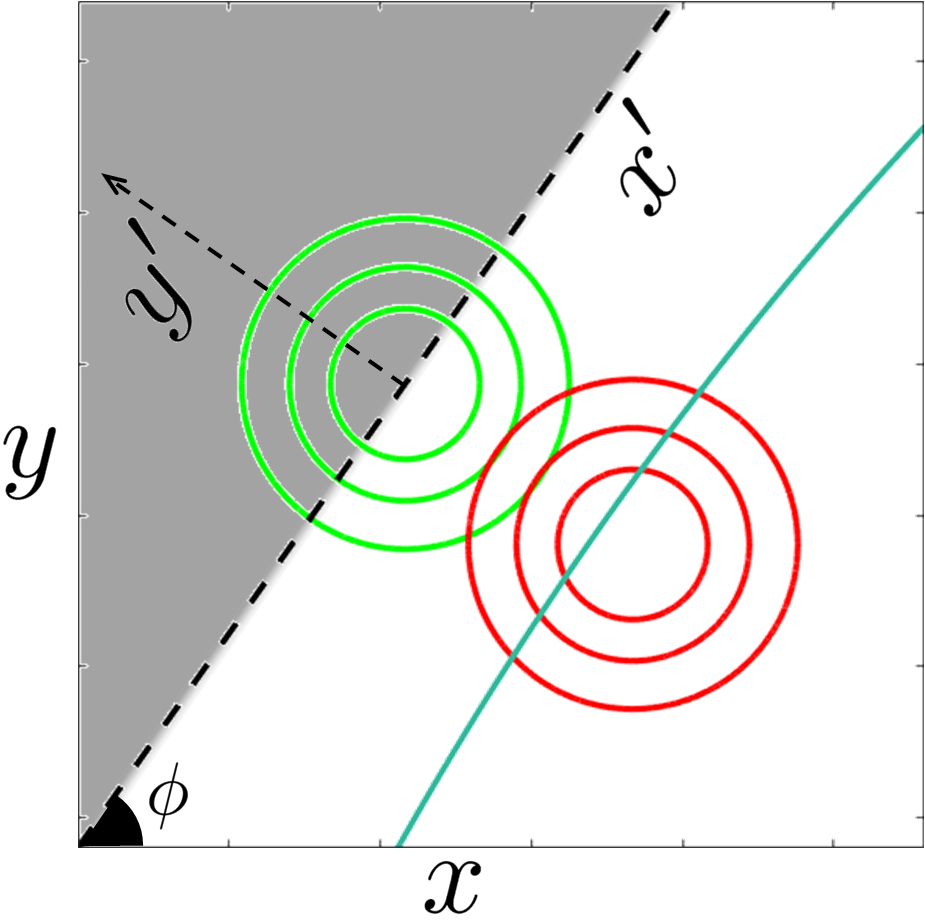}
\caption{Projections in the modified frequency plane $\{x,y\}$ defined in Eq. \eqref{xy} of the probability distributions $P^q(N_s,N_c)$ for the quantum scenario presented in Fig. \ref{Fig2} (red isolines) and $P^c(N_s,N_c)$ for the coherent state defined in \eqref{pscl} (green isoline). The blueish contour line is the estimator given in Eq. \eqref{estimator}. The dashed black line separates the mean values of quantum and classical states as witnessed by a $g^{(2)}(0)$ measurement. In particular, the shaded area includes all states with $g^{(2)}(0) \geq 1$.}
\label{Fig3}
\end{figure}

The question at this stage is what is a good choice for the estimator. %The main objective here is to rule out any classical explanation for the observed statistics, not only the previous one based on the particular coherent state that gives \blue{single events with a frequency $p_s(\rho_{\text{c}})$} \red{the same number of single event on average}, but any model with $p_c\geq p_s^2$. To fulfill this requirement, we wish to build up an estimator that is minimal for the classical state described \blue{in the following} \red{before}, \red{\textit{i.e.}\ for the coherent state reproducing the singles of the quantum scenario}. 
Let us  consider the space of frequencies defined by $(f_s^2,f_c)\equiv\left(\left(\frac{N_s}{N}\right)^2,\frac{N_c}{N}\right).$ We choose a set of coordinates $\{x,y\}$ to cancel the covariance and to equal the variances of $P^q(N_s,N_c)$ in the x and y directions at first order in $\frac{1}{N}$. This is achieved by setting 

\begin{equation}
\label{xy}
\begin{cases}x=\sqrt{\frac{c}{b}}f_s^2+\frac{d}{\sqrt{cb}}f_c\\y=\sqrt{\frac{b}{c}}f_c\end{cases}
\end{equation}
with 
$$\begin{cases}b=\sqrt{\frac{(1-p_{c}(\rho_\text{q}))p_{s}(\rho_\text{q})}{(1-p_{s}(\rho_\text{q}))p_{c}(\rho_\text{q})}-1}\\c=\frac{1-p_{c}(\rho_\text{q})}{2p_{s}(\rho_\text{q})(1-p_{s}(\rho_\text{q}))}\\d=-1\end{cases}.$$ The projection of $P^q(N_s,N_c)$ in the $x-y$ plane hence defines circular isolines, cf. Fig. \ref {Fig3} red isolines. The dashed black line in Fig. \ref {Fig3} distinguishes the frequencies coming from classical and non-classical states. In particular, the distributions with mean values lying on this boundary come from states with $p_c(\rho_{\text{c}}) = p_s^2(\rho_{\text{c}}),$ \textit{i.e.} coherent states with various $p_s.$ The classical scenario that best reproduce the quantum statistics is quite clearly a coherent state which minimizes the Euclidean distance to the quantum distribution, \textit{i.e.} centered on the orthogonal projection of the quantum distribution onto the dashed black line of Fig. \ref{Fig3}. Such a coherent states is associated with 
%In the same way, one can project $P^c(N_s,N_c)$ into the plane $\{x,y\}$ which also defines circular isolines as shown in Fig. \ref{Fig3}, green isolines. \blue{We specify our classical scenario to be rejected by choosing the coherent state which minimizes the Euclidean distance to the quantum distribution in this new coordinate frame. In other words we consider the coherent distribution centered on the orthogonal projection of the quantum distribution onto the dashed black line of Fig. \ref{Fig3}. We naturally assume that this choice is the most apt to reproduce the quantum statistics. This is achieved by setting} 
\begin{equation}
\label{pscl}
p_{s}(\rho_\text{c})=\sqrt{\frac{c(c+d)p_{s}(\rho_\text{q})^2+(d(c+d)+b^2)p_{c}(\rho_\text{q})}{b^2+(c+d)^2}}.
\end{equation}
Calling $(x'_0,y'_0)$ the center of the corresponding distribution $P^c(N_s,N_c),$ an estimator of the form
\begin{equation}
\label{estimator}
\chi =y'-y'_0+a(x'-x'_0)^2,
\end{equation}
where $x'=\cos(\phi)\,x+\sin(\phi)\,y$, $y' =\cos(\phi)\,y-\sin(\phi)\,x$ and $\phi=\arccos{\frac{c+d}{\sqrt{b^2+(c+d)^2}}}$ is intuitively minimized by the coherent state satisfying \eqref{pscl} for appropriate $a$, as $\phi$ is such that the axis of the parabola is orthogonal to the classical/non-classical boundary.\\

\begin{figure}[h]
  \centering
\includegraphics[width=\columnwidth, keepaspectratio=true]{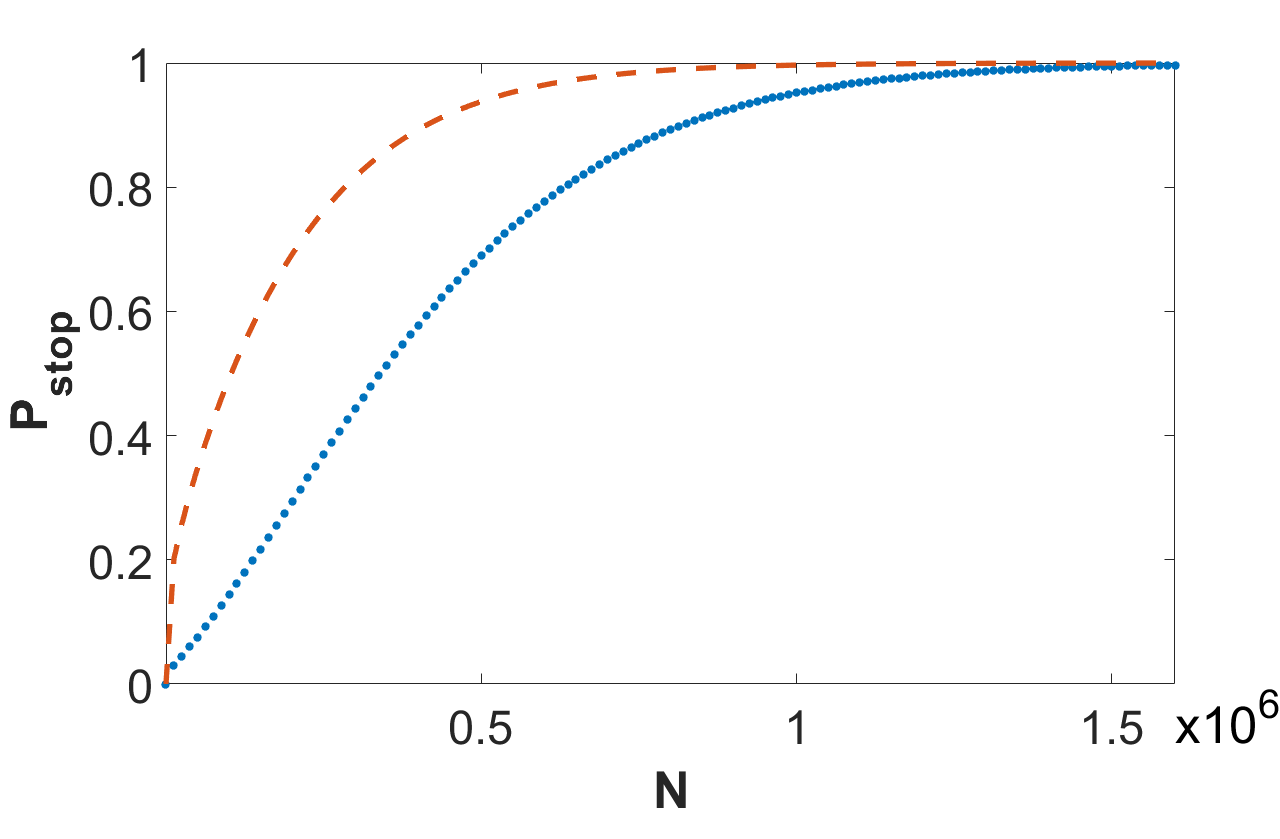}
\caption{Probability to get enough statistics to conclude about non-classicality as a function of the number of runs N for a p-value of $1\%$ (blue dotted line) and $10\%$ (red dashed line).}
\label{Fig4}

%\ref{b} Comparison of the approximated distributions $P(N)$ ($\epsilon=1\%$) for different classical states $\Delta p_s=0$, $\Delta p_s\pm5\times10^{-4}$ (blue line, green dashed line and red dotted line). This shows that the chosen estimator is indeed critical for $\Delta p_s=0,$ \textit{i.e.}\ for a coherent state with $p_s(\rho_\text{c})=p_s(\rho_\text{q})$.}

\end{figure}

The probability that enough statistics is obtained after N runs to exclude the classical distribution $P^c(N_s,N_c)$ with the estimator given in Eq. \eqref{estimator} can be computed numerically as a function of the steepness of the parabola $a$ and the amplitude of displacement operations $\alpha, \beta$. After checking that the considered classical strategy is indeed optimal for the estimator~\eqref{estimator}, we obtained the optimal values $a=40$ and $(\alpha,\beta)\simeq(10.99,0.08)$ for the efficiencies discussed in the previous section and the model of the eye matching the data of Hecht and co-workers $(\theta=7, \eta_e=8\%)$. The results are shown in Fig. \ref{Fig4} for p-values of $1\%$ and $10\%$. We see for example that after $350000$ runs, we have more than $50\%$ chance of being able to rule out classical states with a confidence of $1-\epsilon=99\%$. For $n=12500,$ we find $\langle N\rangle \simeq 402964$ for a confidence of $99\%.$ Note that to perform $403000$ runs with a repetition rate of $1$Hz takes about $112$ hours. The latter provides an upper bound on the timescale of the proposed experiment to get a p-value of $1\%.$ A similar analysis for a p-value of $10\%$ shows that $46$ hours are likely to be enough to detect the non-classical nature of a single photon superposed with vacuum using the human eye. This goes down to 35 hours when considering a threshold at 3 photons while keeping 8\% efficiency and to 29 hours for an efficiency of 10\% and a threshold at 7 photons.

%We foresee for example that $387500$ runs suffice $50\%$ of the time in the case $\epsilon=1\%$. From the derivative of this curve we obtain an approximate probability distribution for the number of runs $N$ which provides an upper bound on the average number of runs $\langle N\rangle$ one would have to perform, cf. Appendix B. This leads to $\langle N\rangle < 439070.$ 

%In Fig. \ref{b}, we compare this probability distribution with the ones obtained after imparting a slight detuning to $p_s(\rho_{\text{c}})$, \textit{i.e.}\ $p_s(\rho_{\text{c}}) = p_s(\rho_{\text{q}})+\Delta p_s$ with $\Delta p_s=\pm5\times10^{-4}$. We observe that the blue line (corresponding to $p_s(\rho_{\text{c}}) = p_s(\rho_{\text{q}})$) is shifted towards larger $N$s even for tiny variations in $p_s$, which in turn validates the statement that our estimator is minimized for a coherent state leading to the same probability of singles than the quantum case. In other words, we can certify the non-classical nature of the measured state by showing that the coherent state leading to $p_s(\rho_{\text{c}}) = p_s(\rho_{\text{q}})$ cannot explain the observed statistics.\\

%Note that to perform $403000$ runs with a repetition rate of $1$Hz takes about $112$ hours. The latter provides an upper bound on the timescale of the proposed experiment to get a p-value of $1\%.$ A similar analysis for a p-value of $10\%$ shows that $46$ hours are \blue{likely to be} enough to detect the non-classical nature of a single photon superposed with vacuum using the human eye. 

\section{Conclusion}
We have presented a concrete proposal for a quantum experiment with the human eye, including the full analysis of the measurement statistics. It uses simple components, namely path-entanglement, displacement operations in phase space and non-photon-number-resolving detectors, to certify with naked eyes the non-classical nature of a state of light. We have given a detailed recipe using parametric conversions and photon-counting techniques only, \textit{i.e.}\ commercially available devices working at room temperature that are routinely used in practice. We have shown that the statistics obtained in a few tens of hours would be sufficient to certify non-classicality with a p-value of $10\%.$ This was obtained with realistic models of the human eye and taking loss and non-unit efficiencies of photon detectors into account. We believe that these timescales are well within reach in practice primarily because the data do not need to be taken in a row. Following in particular the implementation proposed in Fig. \ref{Fig2bis} where a single photon and a coherent state with different polarizations impinge on a polarizing beamsplitter to follow the same optical path and where a set of wave plates and a polarizing beamsplitter are used to make the displacement operations, we can certify from our past experiment \cite{Monteiro15} that the setup is extremely stable even without active stabilization of relative path-length fluctuations. It is thus very likely that the data acquisition can be stopped and started again later for several tens of hours without problem. Despite many preconceptions, we expect the response of the eye to be consistent over long minutes after appropriate dark adaptation. Slow threshold or efficiency drifts can be taken into account easily by periodic re-calibration of the amplitude of displacement operations. We thus see our work as a concrete and realistic proposal to realize the first experiment where the non-classical nature of light is detected directly with the human eye.\\

\section{Acknowledgements} We thank V. Caprara-Vivoli, M. Munsch, Botond Roska, Hendrik Scholl and R. Warburton for valuable discussions. This work was supported by the Swiss National Science Foundation (SNSF), through the NCCR QSIT and the Grant number PP00P2-150579, the John Templeton foundation and the Austrian Science Fund (FWF), Grant number J3462 and P24273-N16.\\

\section{Appendices}

\subsection{Autocorrelation with different arbitrary detectors}

Let us recall the definition of a classical state as given in the main text: $\rho_{cl}=\int \mathrm{d}^2\alpha \-\ p(\alpha)\ketbra{\alpha}{\alpha}$ with $p(\alpha) \geq 0$. We now relax the constraint on the symmetry between the two arms in the autocorrelation measurement, and label $(1,2)$ respectively the reflected and transmitted beams. Each of those beams is sent to a detector which can be different from the other one and the beamsplitter prior to detection is allowed to be unbalanced with coefficients $r/t$. Using the transformation rules for a coherent state on a beamsplitter, it is straightforward to express the probabilities of interest as an integral of the probabilities of singles for appropriate coherent states
\begin{align*}
&P_{s_1}(\rho_{cl})=\int p(\alpha)P_{s_1}\left(\sqrt{r}\alpha\right)\mathrm{d}^2\alpha \\
&P_{s_2}(\rho_{cl})=\int p(\alpha)P_{s_2}\left(\sqrt{t}\alpha\right)\mathrm{d}^2\alpha \\
&P_{c}(\rho_{cl})=\int p(\alpha)P_{s_1}\left(\sqrt{r}\alpha\right)P_{s_2}\left(\sqrt{t}\alpha\right)\mathrm{d}^2\alpha.
\end{align*}
Instead of the autocorrelation which is a ratio of two quantities, we focus on the difference
\begin{align*}
&D(\rho_{cl})=P_{c}(\rho_{cl})-P_{s_1}(\rho_{cl})P_{s_2}(\rho_{cl})\\
&\quad =\int p(\alpha)P_{s_1}\left(\sqrt{r}\alpha\right)P_{s_2}\left(\sqrt{t}\alpha\right)\mathrm{d}^2\alpha\\
&\qquad-\int p(\alpha)P_{s_1}\left(\sqrt{r}\alpha\right)\mathrm{d}^2\alpha\int p(\alpha)P_{s_2}\left(\sqrt{t}\alpha\right)\mathrm{d}^2\alpha,
\end{align*}
Note that $D<0$ implies $g^{(2)}(0)<1.$ Upon inserting $\int p(\beta)\mathrm{d}^2\beta=1$ in $P_c(\rho_{cl})$ and relabelling the dummy variable $\alpha\leftrightarrow\beta$ in some of the terms we get
\begin{align*}
D&(\rho_{cl})=\frac{1}{2}\int\mathrm{d}^2\alpha\-\ p(\alpha)\int\mathrm{d}^2\beta\-\ p(\beta)\\
& \left(P_{s_1}\left(\sqrt{r}\alpha\right)-P_{s_1}\left(\sqrt{r}\beta\right)\right)\left(P_{s_2}\left(\sqrt{t}\alpha\right)-P_{s_2}\left(\sqrt{t}\beta\right)\right)
\end{align*}
We thus obtain that if the functions $P_{s_{1/2}}(\alpha)$ are increasing with $\vert\alpha\vert^2$, then $D(\rho_{cl})\geq0\Leftrightarrow g^{(2)}(0)_{\rho_{cl}}\geq1$, which entails the validity of our witness even in the non-symmetrical case.

\subsection{On the estimation of the average number of runs}
We introduce a formalism to deal with the issue of finding a proper probability distribution for the number of runs. We write the sequence of measurements as a list of zeros and ones, binary stochastic results corresponding respectively to $\chi_{mes}>\chi_{0}(N)$ and $\chi_{mes}\leq\chi_{0}(N)$. It illustrates the situation where an experimenter computes $\chi$ after each measurement (or alternatively after each set of $m$ measurements) and decides if he carries on with the measures (``0") or stops because the results are already satisfactory (``1"). Ideally what we would like to have is the probability $P(n)=P(\underset{n}{\underbrace{0,0,...,1}})$ to reach the required statistics after exactly $n$ runs. Unfortunately, obtaining this ``true" probability numerically represents a computational challenge. What we output from our simulation $P_\text{stop}(N)$ is the probability to get a one at $N^{th}$ position regardless of the preceding sequence. Let's compare the ``cumulative distributions"
\begin{align*}
\sum_{n\leq N}P(n)&=P(1)+P(0,1)+...+P(\underset{N-1}{\underbrace{0,...,0}},1) \\
P(1)&=\sum_{i_{2},...,i_{N}}P(1,i_{2},...,i_{N}) \text{\hspace{3mm}where $i_{k}\in\{0,1\}$}\\
&=P_\text{stop}(N)-\sum_{i_{2},...,i_{N-1}}P(0,i_{2},...,i_{N-1},1)\\
&+\sum_{i_{2},...,i_{N-1}}P(1,i_{2},...,i_{N-1},0)
\end{align*}
\begin{align*}
&P(0,1)-\sum_{i_{2},...,i_{N-1}}P(0,i_{2},...,i_{N-1},1)\\
&=-\sum_{i_{3},...,i_{N-1}}P(0,0,i_{3},...,i_{N-1},1)\\
&+\sum_{i_{3},...,i_{N-1}}P(0,1,i_{3},...,i_{N-1},0)
\end{align*}
\begin{align*}
&\vdots\\
\sum_{n\leq N}P(n)=P_\text{stop}(N)+\sum_{n=0}^{N-2}&P(\underset{n}{\underbrace{0,...,0}},1,i_{n+2},...,i_{N-1},0).
\end{align*}
Therefore $P_\text{stop}(N)\leq \sum_{n\leq N}P(n)$. We would like to translate it into an information on the expectation values. Let us switch to a continuous viewpoint and introduce functions $f$ and $g$ standing for the cumulative distributions, with $\forall x\hspace{1mm} g(x)<f(x)$ (thus $g$ and $f$ replace the $P_\text{stop}$ and $\sum_{n\leq N}P(n)$ of the previous paragraph). We write the expectation values difference and integrate by part
\begin{align*}
\int_{0}^{M}xf'(x)\mathrm{d}x-\int_{0}^{M}xg'(x)\mathrm{d}x&=M[f(M)-g(M)]\\
&-\int_{0}^{M}\underset{>0}{\underbrace{(f(x)-g(x))}}\mathrm{d}x.
\end{align*}
We need to know how the first term behaves when $M\rightarrow\infty$. We haven't find a rigorous way to prove that it vanishes but we notice $M[f(M)-g(M)]<M[1-g(M)]$, which we reasonably assume stays finite based upon the numerical simulations. The latter indeed reveals that $N\longmapsto N\left(1-P_\text{stop}(N)\right)$ shows a decreasing tendency after a given $N$. From this we deduce $\left<N\right>\leq\sum_n n\dfrac{\mathrm{d}P_\text{stop}}{\mathrm{d}n}$.
\bibliographystyle{plain}

\begin{thebibliography}{9}
\bibitem{comment} Recall that any state admits such a representation, however $p(\alpha)$ might not be positive, and hence not a probability distribution.
\end{thebibliography}

\end{document}